\documentclass[a4paper, amsfonts, amssymb, amsmath, onecolumn, showkeys, nofootinbib, twoside]{revtex4-2}
\usepackage{mathtools}
\usepackage{dsfont}

\begin{document}
\title{N-channel parity-time symmetry} 

\author{Ege \"{O}zg\"{u}n}
    \email[Correspondence email address: ]{egeozgun@hacettepe.edu.tr}
    \affiliation{Department of Physics Engineering, Hacettepe University, 06800, Ankara, T\"{u}rkiye}

\begin{abstract} We calculated the eigenvalues for a general N-channel coupled system with parity-time symmetry due to equal loss/gain. We found that the eigenspectrum displays a mixing of parity-time symmetric and broken phases, with $N-2$ of the eigenvalues being parity-time broken whereas the remaining two being either parity-time symmetric or broken depending on the loss/gain and coupling parameters. Our results also show that mixing of parity-time symmetric and parity-time broken phases can only be obtained for at least four-channels if other degrees of freedom like polarization is not taken into account.    

\end{abstract}

\maketitle

\section{Introduction}
Ignited with Bender and Boettcher's seminal paper \cite{Bender}, parity-time $({\cal PT})$ symmetry became a significant theoretical and experimental area of interest. In their work, Bender and Boettcher showed that non-Hermitian Hamiltonians in quantum mechanics can have a partial or a full real eigenspectrum provided that the Hamiltonian is ${\cal PT}$-symmetric. Then in a series of papers, Mostafazadeh generalized the idea and theoretically demonstrated that the Hamiltonians belonging to the class of pseudo-Hermitian Hamiltonians display real spectrum and ${\cal PT}$-symmetric Hamiltonians also belong to that class \cite{M1,M2,M3}. Following the theoretical achievements, ${\cal PT}$ symmetry has found numerous applications, including implementations in optical systems \cite{O1,O2,O3}, waveguides \cite{WG} and single-mode lasers \cite{L}, study of polarization depending scattering in photonic systems \cite{S1} and scattering in spin-$1/2$ systems in quantum mechanics \cite{S2}, implementations in coupled RLC circuits \cite{RLC} and optoelectronic oscillators (OEO)s \cite{OEO1,OEO2}. Recently, a generalized four-channel ${\cal PT}$ symmetry scheme was theoretically suggested for the equal loss/gain setting, that utilizes two different coupling constants \cite{4-ch}. Here we will generalize this scheme two N-channels by assuming the same coupling between all channels and theoretically show that $N-2$ of the eigenvalues of the coupled system are ${\cal PT}$-broken and the remaining two eigenvalues can be either ${\cal PT}$-symmetric or ${\cal PT}$-broken depending on the coupling parameter and the loss/gain value. Thus, for a wide range of parameters the N-channel case displays a mixed ${\cal PT}$ spectrum with coexisting ${\cal PT}$-symmetric and ${\cal PT}$-broken phases.   

The remaining part of the manuscript is organized as the following. In Section \ref{S2} we will go over the simplest scenario i.e., two-channel case as a warm up, then in Section \ref{S3} we will study the four-channel case. In Section \ref{S4} we will show our results for the most general case of N-channels, which is the main result of this manuscript. We finally conclude with Section \ref{S5}.  

\section{Two-channel case} 
\label{S2}
A system consisting of two modes that couple to each other can be expressed in terms of two coupled differential equations, which are known as the coupled-mode equations:

\begin{eqnarray}
    i \dot{a}_1 &=& \omega_1 a_1 -i g a_1 - \kappa_{12} a_2 \nonumber \\
    i \dot{a}_2 &=& \omega_2 a_2 +i l a_2 - \kappa_{21} a_1 
\label{coupled}    
\end{eqnarray}

\begin{figure}[h!]
\centering
\includegraphics[scale=0.7]{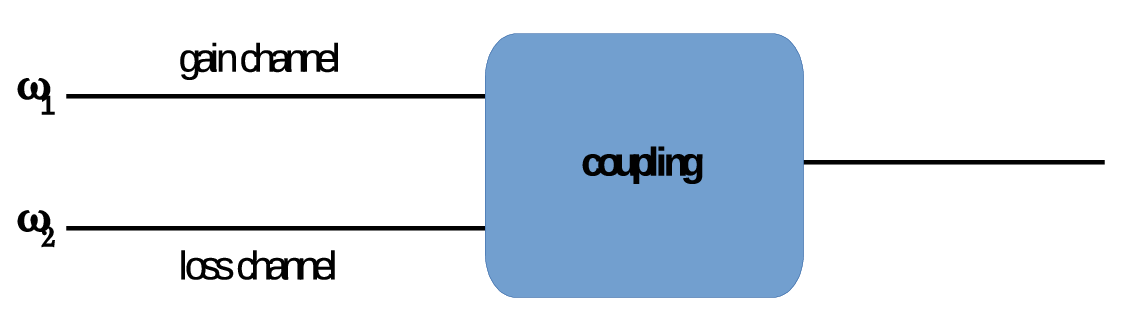}
\caption{A general two-channel scheme. For the equal loss/gain configuration $(g=l)$, ${\cal PT}$ transition can be obtained by tuning either the loss/gain parameter or the coupling constant.}
\label{2-channel}
\end{figure}

In the above equation, $a_i$'s and $\omega_i$'s are the amplitudes and (angular) frequencies of the respective channels $(i=1,2)$, dot denotes time derivative, $\kappa_{12}$ and $\kappa_{21}$ are coupling constants between the channels and $g$ and $l$ are gain and loss parameters, respectively. Figure \ref{2-channel} gives the schematic structure of such two-mode coupling. This formalism is quite general, thus it can be used to describe different systems, such as ${\cal PT}$-symmetric OEOs, coupled waveguides or coupled RLC oscillators. 

From now on, we will assume equal frequencies in all channels $(\omega \coloneqq \omega_1 = \omega_2)$ and equal loss/gain parameters $(\gamma \coloneqq l = g)$. Moreover we will assume real coupling constants, therefore we have $(\kappa \coloneqq \kappa_{12} = \kappa_{21})$. With these assumptions, which can be physically realizable, we will obtain the ${\cal PT}$-symmetric form, that can display ${\cal PT}$-transition for varying coupling constant or loss/gain parameters. The eigenspectrum of the system can be found by solving the characteristic equation for the below matrix obtained from Equation \ref{coupled}:

\begin{eqnarray}
    {\cal M}_2 = 
    \begin{pmatrix}
        \omega+i \gamma & \kappa \\
        \kappa & \omega-i\gamma
    \end{pmatrix}
\end{eqnarray}

By solving  $\det[{\cal M}_2-\mathds{1}_2\lambda]=0$ (where $\mathds{1}_2$ is a $2 \times 2$ identity matrix), we obtain the eigenspectrum of the two-channel system:

\begin{eqnarray}
    \lambda_{1,2} = \omega \pm \sqrt{\kappa^2-\gamma^2}
\end{eqnarray}

Above equation displays the canonical eigenspectrum of a non-Hermitian ${\cal PT}$-symmetric system. ${\cal PT}$ transition from ${\cal PT}$-symmetric phase to ${\cal PT}$-broken phase occurs for $\gamma>\kappa$, where the eigenvalues becomes complex. A significant result worth mentioning is that it is not possible to obtain mixing of ${\cal PT}$-symmetric and ${\cal PT}$-broken phases for the two-channel case (unless other degrees of freedom such as polarization is exploited \cite{S1}). We will show in the following sections that, at least four-channels are required to obtain mixing of ${\cal PT}$-symmetric and ${\cal PT}$-broken phases.

\section{Four-channel case}
\label{S3}
In this Section, we will re-derive the result of Reference \cite{4-ch} but we will take all the coupling constants to be equal which will be useful for the following section when deriving the general result for the N-channel case. For the four-channel case in which channels display equal loss/gain, similar to the two-channel system, we can write four coupled differential equations, which can be then written in the form of a matrix whose characteristic equation gives the four eigenvalues of the system:

\begin{eqnarray}
    {\cal M}_4 = 
    \begin{pmatrix}
        \omega+i \gamma & \kappa & \kappa & \kappa \\
        \kappa & \omega-i\gamma & \kappa & \kappa \\
        \kappa & \kappa & \omega+i \gamma & \kappa \\
        \kappa & \kappa & \kappa & \omega-i\gamma 
    \end{pmatrix}
\end{eqnarray}

Calculating $\det[{\cal M}_4-\mathds{1}_4\lambda]=0$ (where $\mathds{1}_4$ is a $4 \times 4$ identity matrix), yields the eigenspectrum of the four-channel system:

\begin{eqnarray}
    \lambda_{1,2} &=& \omega - \kappa \pm i \gamma \nonumber \\ 
    \lambda_{3,4} &=& \kappa + \omega \pm \sqrt{4 \kappa^2-\gamma^2}
\end{eqnarray}

Above equations display mixing of ${\cal PT}$-symmetric and ${\cal PT}$-broken phases as we mentioned in the previous section. The first two eigenvalues $(\lambda_{1,2})$ are in ${\cal PT}$-broken phase without any dependence on loss/gain and coupling parameters, whereas the last two $(\lambda_{3,4})$ can be either ${\cal PT}$-symmetric or ${\cal PT}$-broken depending on $\kappa$ and $\gamma$. Specifically, for $4 \kappa^2 > \gamma^2$ ${\cal PT}$-symmetric phase is obtained, giving rise to an overall mixing of ${\cal PT}$-symmetric and ${\cal PT}$-broken phases. On the other hand, when $4 \kappa^2 < \gamma^2$ all eigenvalues are in the ${\cal PT}$-broken phase.

\section{N-channel case}
\label{S4}

We are now in a position to study the N-channel system. We limit our interest in the case where $N/2$ of the channels are loss channels whereas the other half are gain channels with equal loss/gain parameters to achieve ${\cal PT}$-symmetry conditions. Moreover, we will assume that all channels couple to each other with the same coupling constant. Same frequencies in all channels are also assumed, similar to the two-channel and four-channel cases. It is important to mention that, in order to obtain  ${\cal PT}$-symmetry conditions $N$ must be even, which should be apparent from the equal loss/gain condition. 

In our theoretical calculations, again we do not make any restrictions on the physical nature of the modes, so the theoretical model is quite general and can be applied to coupled waveguides, coupled RLC-oscillators, OEOs or any system displaying mode-coupling. Therefore, the results we will obtain for the eigenspectrum is going to be valid for any N-channel system respecting the conditions mentioned above for satisfying ${\cal PT}$-symmetry conditions.      

\begin{figure}[h!]
\centering
\includegraphics[scale=0.5]{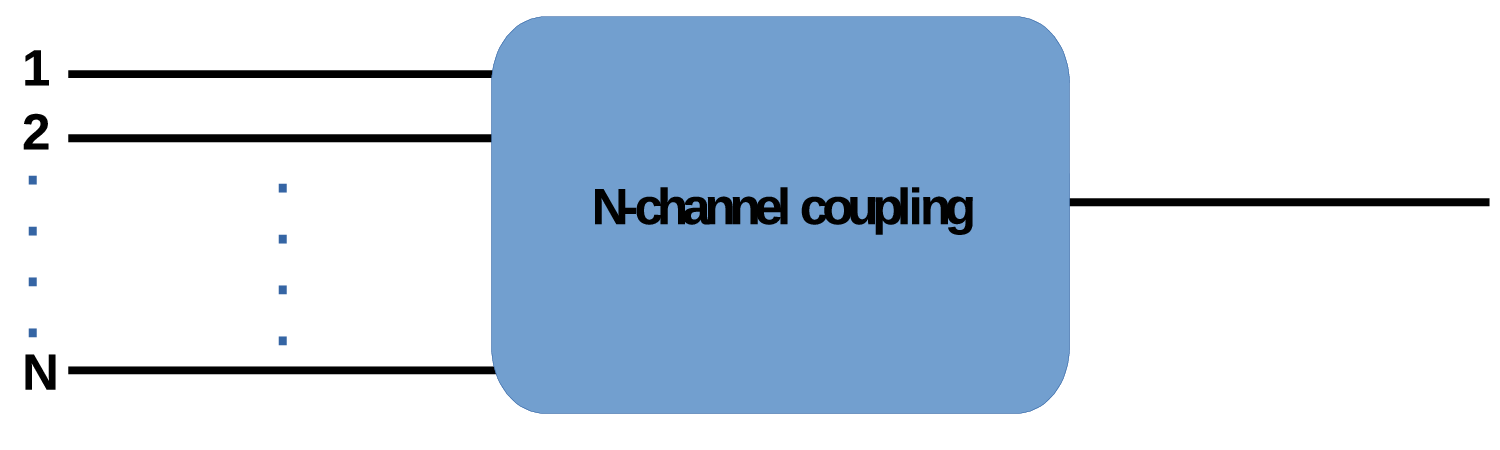}
\caption{Schematic depiction of N-channel coupling. Each of the N modes coming from the left are coupled to other N-1 modes.}
\label{n-channel}
\end{figure}

Figure \ref{n-channel} displays the schematic demonstration of the N-channel system. We will again write the coupled-mode equations in the matrix form. For the N-channel case we will have an $N \times N$ matrix, which will yield $N$ eigenvalues from its characteristic equation. The form of the matrix for the N-channel case is given below:

\begin{eqnarray}
{\cal M}_N =
    \begin{pmatrix}
        \omega+i\gamma & \kappa & \ldots & \ldots & \kappa \\
        \kappa & \omega-i\gamma & \kappa & \ldots & \vdots \\
        \vdots & \kappa & \ddots &  \ddots & \vdots \\
        \vdots & \ldots & \ddots & \ddots & \kappa \\
        \kappa & \ldots  & \ldots &  \kappa & \omega-i\gamma             
        
    \end{pmatrix}
\end{eqnarray}
The eigenvalues can again be found by solving $\det[{\cal M}_N-\mathds{1}_N\lambda]=0$ (where $\mathds{1}_N$ is an $N \times N$ identity matrix) which can be calculated analytically from the equation for $\lambda$ given below:

\begin{eqnarray}
    & \Big [&\lambda-(\omega-\kappa+i\gamma) \Big ]^{\left (\frac{N}{2}-1 \right)} \Big [\lambda-(\omega-\kappa-i\gamma) \Big ]^{ \left (\frac{N}{2}-1 \right )} \nonumber \\ \times & \Bigg [ &\lambda^2-2 \left (\omega+\left [\frac{N}{2}-1 \right ]\kappa \right )\lambda+ \left (\omega+\left [ \frac{N}{2}-1 \right ]\kappa \right )^2+\gamma^2-\left ( \frac{N}{2} \right ) ^2 \kappa^2 \Bigg ] = 0
\label{char}
\end{eqnarray}

The above equation is an Nth order polynomial equation for $\lambda$ with its solutions yielding the $N$ eigenvalues of the system:

\begin{eqnarray}
    \lambda_1 = \lambda_2 = \ldots = \lambda_{(N/2-1)} = \omega - \kappa +i\gamma \nonumber \\
    \lambda_{(N/2)} = \lambda_{(N/2+1)} = \ldots = \lambda_{(N-2)} = \omega - \kappa -i\gamma \nonumber \\
    \lambda_{(N-1)}= \left [ \frac{N}{2}-1 \right ] \kappa + \omega + \sqrt{\left [ \frac{N\kappa}{2} \right]^2-\gamma^2} \nonumber \\ 
    \lambda_{(N)}= \left [ \frac{N}{2}-1 \right ] \kappa + \omega - \sqrt{\left [ \frac{N\kappa}{2} \right]^2-\gamma^2}
\label{eigen}
\end{eqnarray}

\begin{figure}[h!]
\centering
\includegraphics[scale=1]{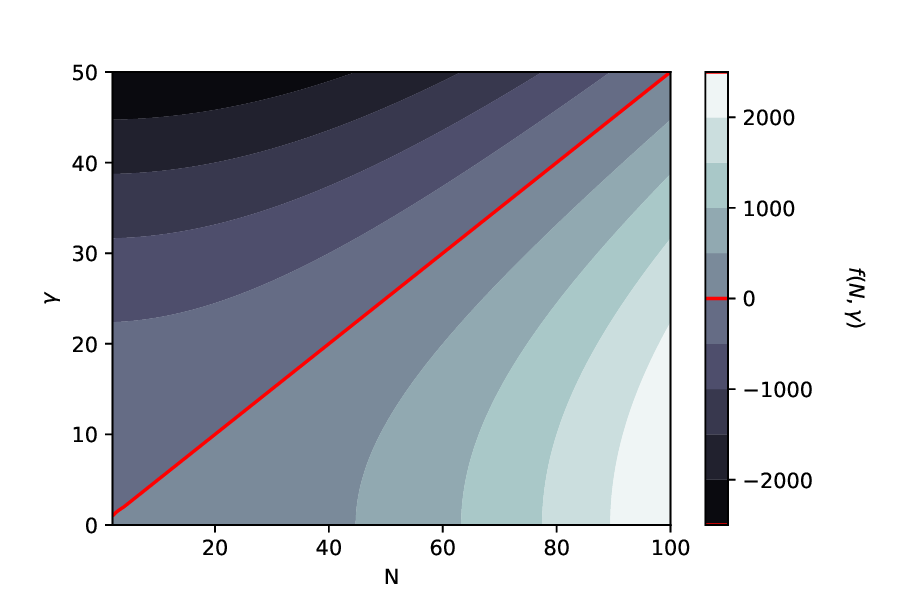}
\caption{Counter plot of the function $f(N,\gamma) = [N/2]^2-\gamma^2$ for a wide range of the parameters $N$ and $\gamma$. The red line separates the ${\cal PT}$-symmetric and ${\cal PT}$-broken regions for the eigenvalues $\lambda_{(N-1)}$ and $\lambda_{(N)}$ where the lighter region corresponds to ${\cal PT}$-symmetric phase and the darker region corresponds to ${\cal PT}$-broken phase.}
\label{pt-reg}
\end{figure}

As it can be seen from Equation \ref{eigen}, $N-2$ of the eigenvalues, consisting of two pairs with $(N-2)/2$-fold degeneracy are in ${\cal PT}$-broken phase. The remaining two eigenvalues can be either both ${\cal PT}$-symmetric or ${\cal PT}$-broken depending on the values of loss/gain and coupling parameters. By observing Equation \ref{char}, we can also arrive to the same conclusion which we mentioned in Section \ref{S2} that for mixed ${\cal PT}$-symmetric and ${\cal PT}$-broken phases, we need at least four-channels unless other degrees of freedom such as polarization are taken into account.    

Let us now define a function for fixed coupling constant, say $\kappa=1$ to illustrate the ${\cal PT}$-symmetric and  ${\cal PT}$-broken phases of $ \lambda_{(N-1)}$ and $ \lambda_{(N)}$ for different values of $N$ and $\gamma$:

\begin{eqnarray}
f(N,\gamma) = [N/2]^2-\gamma^2.    
\end{eqnarray}

The function $f(N,\gamma)$ has two distinct regions. When $f(N,\gamma)<0$ the two eigenvalues $\lambda_{(N-1)}$ and $\lambda_{(N)}$ are in the ${\cal PT}$-broken phase, whereas for $f(N,\gamma)>0$ they are ${\cal PT}$-symmetric, which demonstrates an overall mixing of the eigenspectrum with coexisting ${\cal PT}$-symmetric and ${\cal PT}$-broken phases. We plot $f(N,\gamma)$ for a wide range of the parameters $N$ and $\gamma$ in Figure \ref{pt-reg}. It can be seen that as the number of channels N increases, ${\cal PT}$-symmetric phase becomes more and more robust for fixed loss/gain value.

\section{Conclusion}
\label{S5}

We calculated the eigenspectrum of an N-channel system consisting of even number of channels with $N/2$ pairs of them having equal loss/gain and showed that the spectrum consists of two pairs of ${\cal PT}$-broken eigenvalues each with a degeneracy of $(N-2)/2$ and two other eigenvalues that can either be in the ${\cal PT}$-symmetric or ${\cal PT}$-broken phase depending on the values of the loss/gain parameters and the coupling constant. Another significant result we obtained is the lower limit for obtaining mixed ${\cal PT}$-symmetric and ${\cal PT}$-broken phases, which is at least four-channels, assuming no other degrees of freedom such as polarization is exploited. Moreover, our results showed that, for the increasing number of channels, the ${\cal PT}$-symmetric phase becomes more robust for fixed loss/gain values. 

\section*{Acknowledgment}
The author acknowledges fruitful discussions with G\"{o}khan Alka\c{c}.


\begin{thebibliography}{99}
\bibitem{Bender} C. M. Bender and S Boettcher, Real Spectra in Non-Hermitian Hamiltonians Having PT Symmetry, Physical Review Letters, 80, 5243 (1998). 

\bibitem{M1} A.  Mostafazadeh, Pseudo-Hermiticity versus PT-symmetry: The necessary condition for the reality of the spectrum of a non-Hermitian Hamiltonian, Journal of Mathematical Physics 43, 205 (2002). 

\bibitem{M2} A.  Mostafazadeh, Pseudo-Hermiticity versus PT-symmetry. II. A complete
characterization of non-Hermitian Hamiltonians with a real spectrum, Journal of Mathematical Physics 43, 2814 (2002). 

\bibitem{M3} A.  Mostafazadeh, Pseudo-Hermiticity versus PT-symmetry III: Equivalence of pseudo-Hermiticity and the presence of antilinear symmetries, Journal of Mathematical Physics 43, 3944 (2002). 

\bibitem{O1} R. El-Ganainy, K. G. Makris, D. N. Christodoulides, and Ziad H. Musslimani, Theory of coupled optical PT-symmetric structures, Optics Letters 32, 2632 (2007).


\bibitem{O2} K. G. Makris R. El-Ganainy and D. N. Christodoulides, Beam Dynamics in PT Symmetric Optical Lattices, Physical Review Letters, 100, 103904 (2008). 

\bibitem{O3} Christian E. R\"{u}ter, K. G. Makris, Ramy El-Ganainy, D. N. Christodoulides,
Mordechai Segev and Detlef Kip, Observation of parity–time symmetry in optics, Nature Physics, 6, 192 (2010).

\bibitem{WG} Shachar Klaiman, Uwe G\"{u}nther, and Nimrod Moiseyev, Visualization of Branch Points in PT-Symmetric Waveguides, Physical Review Letters, 101, 080402 (2008). 

\bibitem{L} L. Feng, Z. J. Wong, R.-M. Ma, Y. Wang, X. Zhang, Single-mode laser by parity-time
symmetry breaking, Science, 346, 972 (2014).

\bibitem{S1} E. \"{O}zg\"{u}n, A. E. Serebryannikov, E. Ozbay and C. M. Soukoulis, Broadband mixing of PT-symmetric and PT-broken phases in photonic heterostructures with a one-dimensional loss/gain bilayer, Scientific Reports, 7, 15504 (2017). 

\bibitem{S2} E. \"{O}zg\"{u}n, T. Hakio\u{g}lu and E. Ozbay, Scattering of spin-1/2 particles from a PT-symmetric complex potential, Europhysics Letters, 131, 11001 (2020).

\bibitem{RLC} Y. Choi, C. Hahn, J. W. Yoon and S. H. Song, Observation of an anti-PT-symmetric exceptional
point and energy-difference conserving dynamics in electrical circuit resonators, Nature Communications, 9, 2182 (2018).

\bibitem{OEO1} J. Zhang and J. Yao, Parity-time–symmetric optoelectronic oscillator, Science Advances, 4, eaar678 (2018). 

\bibitem{OEO2} E. \"{O}zg\"{u}n, F. Uyar, T. Kartaloglu, E. Ozbay and I. Ozdur, A parity-time-symmetric optoelectronic oscillator with polarization multiplexed channels, Journal of Optics, 24, 055802 (2022).
 
\bibitem{4-ch} E. \"{O}zg\"{u}n, E. Ozbay and I. Ozdur, Four-channel parity-time symmetry, Europhysics Letters, 140, 10001 (2022).

\end{thebibliography}
\end{document}